\documentclass[12pt]{article}
\usepackage{amsmath,amssymb,bm,graphicx}
\usepackage{cite}

\title{Tunneling Mechanism in Kerr-Newman Black Hole\\ and Dimensional Reduction near the Horizon}
\date{}
\makeatletter
 
  \@addtoreset{equation}{section}
 \makeatother 
\begin{document}
\begin{flushright}
MISC-2010-06
\end{flushright}
\vspace{0.6cm}

\begin{center}
{\large{\bf 
Tunneling Mechanism in Kerr-Newman Black Hole\\ and Dimensional Reduction near the Horizon}}
\end{center}
\vspace{1cm}

\begin{center}
Koichiro Umetsu\footnote{E-mail: umetsu@cc.kyoto-su.ac.jp}
\end{center}
\begin{center}
{\it Maskawa Institute for Science and Culture, Kyoto Sangyo University,\\ Kyoto, 603-8555, Japan}
\end{center}

\vspace{1cm}

\abstract{
It is shown that the derivation of the Hawking radiation from a rotating black hole on the basis of the tunneling mechanism
is greatly simplified by using the technique of the dimensional reduction near the horizon.
This technique is illustrated for the original derivation by Parikh and Wilczek, but it is readily applied to a variant of the method such as suggested by Banerjee and Majhi.
}
\newpage
\section{Introduction}

The black hole radiation is explained by taking into account the quantum effect in the framework of general relativity.
This is commonly called Hawking radiation.
Hawking radiation is one of very important phenomena where both of general relativity and quantum theory play a role.
After Hawking's original derivation \cite{haw01}, various derivations of Hawking radiation have been suggested.

A method of deriving Hawking radiation based on quantum tunneling was proposed by Parikh and Wilczek \cite{par01}.
The essential idea is as follows.
We consider that a particle-antiparticle pair is formed close to the horizon inside a black hole.
We can divide the field associated with a particle into ingoing modes moving toward the center of the black hole
and outgoing modes trying to move outside the black hole.
The ingoing modes are trapped inside the horizon.
On the other hand, although the horizon plays a role of a barrier when the outgoing modes try to move outside the black hole, 
a part of the outgoing modes move outside the black hole by the quantum tunneling effect.
If the particle which comes out to our universe has positive energy, such a particle (and also an antiparticle) can stably exist 
and we can regard the particle outside the horizon as the radiation from the black hole.
Parikh and Wilczek calculated the WKB probability amplitude for the particle by taking into account classically forbidden paths.
By comparing the probability with the Boltzmann factor in thermodynamics, they successfully derived the Hawking temperature.
The physical picture of the derivation of Parikh and Wilczek is quite clear.

They derived these results for spherically symmetric black holes (in particular, a Schwarzschild black hole and a Reissner-Nordstr\"om black hole).
The extension of their derivation to rotating black holes, which have spherically asymmetric geometry 
such as Kerr and Kerr-Newman black holes, has been discussed in the literature \cite{zha01,zha02,jia01,che01}.
The derivations for rotating black holes are mathematically quite involved because of the effects of the rotation.

In this paper, we show a new approach which extends the method of Parikh and Wilczek to a rotating black hole.
It is shown that the derivation of the Hawking temperature is greatly simplified 
if one uses the technique of the dimensional reduction near the horizon.

The essential idea of the dimensional reduction is as follows: We consider the action for a scalar field in the Kerr-Newman black hole background.
We expand the scalar field in terms of the spherical harmonics and use the tortoise coordinate \cite{whe01,reg01} to find the behavior of the field near the horizon.
We can then ignore the mass, potential and interaction terms in the action because the kinetic term dominates in the high-energy theory near the horizon. 
We thus find that the integrand in the action does not depend on angular variables
when only the dominant terms near the horizon are retained.
The 4-dimensional action is then reduced to a 2-dimensional action by using the orthonormal condition of the spherical harmonics.
As a result, we can show that the 4-dimensional spherically asymmetric Kerr-Newman black hole effectively behaves as 
a 2-dimensional spherically symmetric black hole by this technique of the dimensional reduction.

We note that this technique is valid only for the region very close to the horizon. 
The use of the technique in the tunneling mechanism is justified since the tunneling effect is the
quantum effect arising within the Planck length near the horizon region.
Our approach greatly simplifies the derivation of the Hawking temperature from a rotating black hole in comparison with previous works \cite{zha01,zha02,jia01,che01}.
This technique of dimensional reduction have been used in the derivation of Hawking radiation on the basis of quantum anomalies \cite{iso01,iso02,mur01,ume01}.
We have shown elsewhere \cite{ume02} that this dimensional reduction technique is useful in the analysis of a rotating black hole in the modified tunneling scheme by Banerjee and Majhi \cite{ban02}.
More recently, it was discussed by Banerjee and Majhi that the technique is useful when the unified description of Hawking and Unruh effects
is extended to a rotating black hole \cite{ban01}.
To my knowledge, the usefulness of the dimensional reduction has not been discussed in the framework of the tunneling method of Parikh and Wilczek. 

The contents of this paper are follows.
In Section 2, we show how to extend the method of Parikh and Wilczek to the Kerr-Newman black hole by using the technique of the dimensional reduction near the horizon and how to derive the Hawking temperature.
Section 3 is devoted to discussion and conclusion.

\section{Extension to Kerr-Newman black hole}

Parikh and Wilczek derived Hawking radiation on the basis of the quantum tunneling effect in the 4-dimensional spherically symmetric black hole backgrounds.
For these black holes, the line element is given by
\begin{align}
ds^2=-f(r)dt^2_{{\rm s(r)}}+\frac{1}{f(r)}dr^2+r^2d\theta^2 +r^2 \sin \theta d\varphi^2,
\label{sph01}
\end{align}
where $f(r)$ is given by $f(r)=1-\frac{2M}{r}$ for the Schwarzschild black hole with the mass $M$ and $f(r)=1-\frac{2M}{r}+\frac{Q^2}{r^2}$ for the Reissner-Nordstr\"om black hole with the mass $M$ and charge $Q$.

We now consider the case of a charged and rotating Kerr-Newman black hole.
The line element is given by
\begin{align}
ds^2=&-\frac{\Delta -a^2 \sin^2\theta}{\Sigma}dt_{{\rm k}}^2-\frac{2a\sin^2\theta}{\Sigma}(r^2+a^2-\Delta)dt_{{\rm k}}d\varphi \notag\\
&-\frac{a^2\Delta \sin^2\theta-(r^2+a^2)^2}{\Sigma}\sin^2\theta d\varphi^2+\frac{\Sigma}{\Delta}dr^2+\Sigma d\theta^2,
\label{kn01}
\end{align}
where the notations are respectively defined by
\begin{align}
a&\equiv \frac{J}{M},
\label{kn02}\\
\Sigma&\equiv r^2+a^2\cos^2 \theta,
\label{kn03}\\
\Delta&\equiv r^2-2Mr+a^2+Q^2=(r-r_+)(r-r_-).
\label{kn04}
\end{align}
Here, $M$, $J$, $Q$ and $r_{+(-)}$ are respectively a mass, an angular momentum, an electrical charge and the outer (inner) horizon of the Kerr-Newman black hole defined by
\begin{align}
r_\pm=M\pm \sqrt{M^2-a^2-Q^2}.
\label{kn05}
\end{align}
It follows from the expression (\ref{kn01}) that the Kerr-Newman metric has spherically asymmetric geometry.

We show that the technique of the dimensional reduction near the horizon is very useful,
when we extend the method of Parikh and Wilczek to the Kerr-Newman black hole.
By using the technique, 
it is shown that the Kerr-Newman line element (\ref{kn01}) behaves as the 2-dimensional spherically symmetric line element
\begin{align}
ds^2=-f(r)dt_{{\rm k}}^2+\frac{1}{f(r)}dr^2,
\label{kerr5}
\end{align}
where $f(r)$ is given by
\begin{align}
f(r)\equiv \frac{\Delta}{r^2+a^2} =1-\frac{2Mr}{r^2+a^2} +\frac{Q^2}{r^2+a^2},
\end{align}
and contains the effect of the rotation $a$.
For details of this calculation, see \S 2 in \cite{ume02} or \S 2.4 in \cite{ume03}.
This technique of the dimensional reduction is valid only for the region very close to the horizon.
The use of the same technique in the tunneling method of Parikh and Wilczek is justified 
because the tunneling method deals with the region very close to the horizon. 

To describe the across-horizon phenomena, Parikh and Wilczek used the Painlev\'e coordinates \cite{pai01}.
Thus we define the time component of the Painlev\'e-like coordinates corresponding to (\ref{kerr5}) by
\begin{align}
dt\equiv dt_{{\rm k}}+ \frac{\sqrt{\frac{2Mr}{r^2+a^2} - \frac{Q^2}{r^2+a^2}}}{1-\frac{2Mr}{r^2+a^2} + \frac{Q^2}{r^2+a^2}}dr.
\end{align}
The Painlev\'e-like line element is then given by
\begin{align}
ds^2=-\left( 1-\frac{2Mr}{r^2+a^2} + \frac{Q^2}{r^2+a^2}\right)dt^2 + 2\sqrt{\frac{2Mr}{r^2+a^2} - \frac{Q^2}{r^2+a^2}}dtdr +dr^2.
\label{met01}
\end{align}
The radial null geodesics is also given by
\begin{align}
\dot{r} \equiv \frac{dr}{dt}=\pm 1 -\sqrt{\frac{2Mr}{r^2+a^2} - \frac{Q^2}{r^2+a^2}},
\label{null01}
\end{align}
where the sign $+$ $(-)$ stands for the outgoing (ingoing) geodesic.

When we take into account the effect of the particle's self-gravitation \cite{kra01},
we replace $M$ by $M-\omega$, where $\omega$ is the energy of the particle which escapes from the black hole by the tunneling mechanism.
Thus, (\ref{met01}) and (\ref{null01}) are respectively rewritten as
\begin{align}
ds^2&=-\left( 1-\frac{2(M-\omega)r}{r^2+a^2} + \frac{Q^2}{r^2+a^2}\right)dt^2 + 2\sqrt{\frac{2(M-\omega)r}{r^2+a^2} - \frac{Q^2}{r^2+a^2}}dtdr +dr^2,
\label{met02}
\end{align}
\begin{align}
\dot{r} =\pm 1 -\sqrt{\frac{2(M-\omega)r}{r^2+a^2} - \frac{Q^2}{r^2+a^2}}.
\label{null02}
\end{align}

By following Parikh and Wilczek, we evaluate the WKB probability amplitude for a classically forbidden trajectory.
The imaginary part of the action for an outgoing positive energy particle, which crosses the horizon outwards from $r_{{\rm in}}$ to $r_{{\rm out}}$, is given by
\begin{align}
{\rm Im}~ S={\rm Im}\int^{r_{{\rm out}}}_{r_{{\rm in}}}p_r dr = {\rm Im}\int^{r_{{\rm out}}}_{r_{{\rm in}}} \int^{p_r}_{0} dp'_r dr.
\label{ims01}
\end{align}
By using Hamilton's equation $\dot{r}=+\frac{dH}{dp_r}\Big |_r$ in (\ref{ims01}), we obtain
\begin{align}
{\rm Im}~ S={\rm Im} \int^{M-\omega}_{M} \int^{r_{{\rm out}}}_{r_{{\rm in}}}\frac{dr}{\dot{r}} dH
= {\rm Im}\int^{+\omega}_{0} \int^{r_{{\rm out}}}_{r_{{\rm in}}}\frac{dr}{1 -\sqrt{\frac{2(M-\omega')r}{r^2+a^2}- \frac{Q^2}{r^2+a^2}}} (-d\omega'),
\label{ims02}
\end{align}
where we used $H=M-\omega'$.
By using Feynman's $i\epsilon$ prescription $\omega \to \omega -i\epsilon $, 
we have
\begin{align}
{\rm Im}~ S=+{\rm Im} \int^{r_{{\rm out}}}_{r_{{\rm in}}}\int^{M-\omega}_{M} \frac{dM'}{1 -\sqrt{\frac{2M'r}{r^2+a^2}- \frac{Q^2}{r^2+a^2}}} dr=-\pi\int^{r_{{\rm out}}}_{r_{{\rm in}}} \left( \frac{r^2+a^2}{r} \right) dr,
\label{ims03}
\end{align}
where $r_{{\rm out}}$ and $r_{{\rm in}}$ are respectively given by
\begin{align}
r_{{\rm out}}&=M-\omega + \sqrt{(M-\omega)^2 -a^2 -Q^2},\\
r_{{\rm in}}&=M+\sqrt{M^2-a^2-Q^2}=r_+.
\end{align}
When we consider the effect in the first order of $\omega$, (\ref{ims03}) becomes 
\begin{align}
{\rm Im}~ S&=-\frac{\pi}{2}\left( r^2_{{\rm out}} -r^2_{{\rm in}}\right)-\pi a^2 \ln \left( \frac{r_{{\rm out}}}{r_{{\rm in}}} \right)\\
&\simeq \frac{\pi(r_+^2+a^2)}{r_+-M}\omega,
\end{align}
where we used 
\begin{align}
r_{{\rm out}}\simeq r_+ - \frac{r_+}{r_+ -M}\omega.
\end{align}
Thus, we obtain the WKB probability amplitude
\begin{align}
\Gamma = e^{- 2 {\rm Im}~ S} \simeq e^{-\frac{2\pi}{\kappa}\omega},
\label{gam01}
\end{align}
where we used $\kappa$ which is the surface gravity on the horizon of the Kerr-Newman black hole defined by
\begin{align}
\kappa\equiv \frac{1}{2} \partial_r f(r)\Big|_{r=r_+}=\frac{r_+-M}{r_+^2 +a^2}.
\end{align}
By comparing the result (\ref{gam01}) with the Boltzmann factor ($\Gamma=e^{-\frac{\omega}{T}}$) in a thermal equilibrium state at temperature $T$,
we find the Hawking temperature $T_{{\rm H}}$
\begin{align}
T_{{\rm H}}=\frac{\kappa}{2\pi},
\end{align}
which is the desired result \cite{jia01}.

\section{Discussion and Conclusion}
We have shown that we can extend the method of Parikh and Wilczek to the case of the Kerr-Newman black hole by using the technique of the dimensional reduction near the horizon and we can derive the Hawking temperature in a very simple manner.
As already stated in Introduction, the extension of the tunneling method of Parikh and Wilczek to the rotating black hole had been previously discussed in \cite{par01, zha01, zha02, jia01, che01}.
However, their analyses are much more involved in comparison with the method shown in this paper
because both the effects of $p_\varphi$, which is the canonical momentum conjugate to $\varphi$, and $\dot{\varphi}$, which is the time derivative of $\varphi$, have to be analyzed in their method.

The calculation in our derivation is little different from the original calculation for spherically symmetric black holes by Parikh and Wilczek.
Nevertheless, our method includes the essential effect of the rotation, and the Hawking temperature is correctly derived by evaluating 
the effect to the first order in $\omega$.
By using the present technique, we can also understand the reason why Parikh and Wilczek consider only the $(t-r)$ sector of the line element when they derive the radial null geodesics (\ref{null01}).

We thus conclude that the technique of the dimensional reduction is very useful in the derivation of Hawking radiation on the basis of the tunneling mechanism.
Recently, a variant of the tunneling method of Parikh and Wilczek was proposed by Banerjee and Majhi \cite{ban02}.
It was shown elsewhere that the technique of the dimensional reduction is also useful when their method is extended to analyze the tunneling effect in a rotating black hole background \cite{ume02}.

In passing, we note that the authors in \cite{akh04} discuss the potential ``factor of two subtlety'' in the tunneling method, but the main issue in the present paper is away from the subtlety.

\section*{Acknowledgements}

I would like to thank Prof. K. Fujikawa for helpful discussions and for a careful reading of the manuscript.
I also appreciate Prof. D. Singleton for a useful correspondence.
This work is supported by Maskawa Institute in Kyoto Sangyo University.


\end{document}